\documentclass[12pt,a4paper]{article}

\usepackage{amsmath}
\usepackage{amssymb}
\usepackage{mathrsfs}
\usepackage{graphicx,subcaption}
\usepackage{dsfont}
\usepackage[colorlinks=true,linkcolor=black,citecolor=black,urlcolor=black]{hyperref}
\usepackage{enumerate}
\usepackage{epstopdf}

\numberwithin{equation}{section}

\def\eqn#1{ \begin{eqnarray} #1 \end{eqnarray} }
\def\eq#1 { \begin{equation} #1 \end{equation} }
\def\eqn#1{ \begin{eqnarray} #1 \end{eqnarray} }

\def\w{\omega}
\def\k{\kappa}

\def\d{\partial}

\def\cO{\mathcal{O}}

\def\sl2r{SL(2,\mathbb{R})}
\def\zb{{\bar z}}
\def\gt{{\tilde g}}

\def\ce{\varepsilon}

\newcommand{\lsim}{\mathrel{\hbox{\rlap{\lower.55ex \hbox{$\sim$}} \kern-.3em \raise.4ex \hbox{$<$}}}}
\newcommand{\gsim}{\mathrel{\hbox{\rlap{\lower.55ex \hbox{$\sim$}} \kern-.3em \raise.4ex \hbox{$>$}}}}

 \newcommand{\be}{\begin{equation}}
\newcommand{\ee}{\end{equation}}

\newcommand{\vev}[1]{\left< #1 \right>} 

\usepackage{bm}

\begin{document}

\title{\begin{flushright}\vspace{-1in}
       \mbox{\normalsize  EFI-15-10}
       \end{flushright}
       \vskip 20pt
Viscosities and shift in a chiral superfluid: \\ a holographic study}

\date{\today}

\author{
Siavash Golkar
\thanks{\href{mailto:golkar@uchicago.edu}         {golkar@uchicago.edu}} ,
   Matthew M. Roberts
   \thanks{\href{mailto:matthewroberts@uchicago.edu}
     {matthewroberts@uchicago.edu}}~
      \\ \\
   {\it \it Kadanoff Center for Theoretical Physics and Enrico Fermi Institute,}\\
   {\it   University of Chicago, 5640 South Ellis Ave., Chicago, IL 60637 USA}
} 

\maketitle

\begin{abstract}
We consider a holographic model of chiral superfluidity whose bulk is Einstein Yang-Mills and compute viscosity and conductivity responses  away from the probe limit. We calculate Hall viscosity and analyze its relationship to the superfluid density and the shift. We find that the relationship between these quantities derived from effective field theory at zero temperature persists for all temperatures: for $p\pm ip$ their ratio is equal to $\mp1/2$. At low temperatures the system develops a Lifshitz throat, indicating an anisotropic scaling symmetry in the infrared dynamics.\end{abstract}

\newpage

\section{Introduction}
\label{sec:intro}
The promise of providing a framework to construct Majoranna fermions and its potential application to fault tolerant quantum computing has made high temperature chiral superfluids an attractive subject of study in recent years \cite{Moore:1991ks,Kitaev-QC,Nayak-QC}. While traditional superconductivity is well understood via the BCS pairing mechanism, high temperature superconductivity cannot be explained by conventional electron-phonon interactions that dominate the pairing mechanism in BCS.

There has been much advancement in the understanding of these systems \cite{PhysRevB.77.144516,PhysRevB.77.174513,alicea2012new}, however a universal understanding which can explain different physical aspects is not currently in hand. As an example, the stress response  at finite temperature has not been fully explored. Given the possible practical application of these systems in quantum computing, it is important that their tensile properties be fully analyzed and understood. 

One possible venue for arriving at universal predictions is by considering the effective field theory (EFT) that governs the dynamics at very low temperature and energies. The main guiding principle of this approach is the underlying symmetries of the system. For example, if we consider consider a superfluid system controlled mainly by a quantum critical point with Lorentz invariance, the EFT must also be Lorentz invariant. These symmetries are often not powerful enough to completely constrain the low energy dynamics of the system, however, in some cases when combined with topological data they can provide predictions that are either universal or can relate various transport coefficients. This approach was applied to superfluid systems at zero temperature in \cite{Hoyos:2013eha,Golkar:2014paa,Moroz:2014ska}. In particular it was shown that the ratio of hall viscosity to the superfluid density can be determined by topological considerations, i.e. it is given by the superfluid shift, which relates the ground state vortex number of the system to the spatial topology. Specifically, the ratio of Hall viscosity to superfluid density is  equal to $\mp 1/2$ for $p\pm ip$  superfluid. However, by their very essence, these analyses are not expected to be valid at finite temperature and since they rely on an action principle, they also cannot describe any dissipative physics. 

Traditionally, in areas of condensed matter physics where a careful first principle analysis is not available or difficult, holography has stepped in to provide valuable intuition by providing access to a very different set of universality classes than that of weakly interacting quasiparticles. Since these systems are the result of deforming conformal theories, the inherited conformal symmetry further constrains the transport properties of the system.

In this paper we use a holographic model to compute the stress response of relativistic chiral superfluids. The model we consider is Einstein-Yang-Mills with group $SU(2)$ \cite{Gubser:2008zu,Roberts:2008ns}, where, after breaking $SU(2)\rightarrow U(1)$ the W-bosons in the bulk provide a charged vector operator in the boundary. Studying stress response requires looking at the system away from the probe limit (where the charged and neutral sectors decouple) and so we must allow for backreaction. We also  analyze the electronic responses that were previously computed in the probe limit and discuss how they change. The main result of this analysis is that the ratio of Hall viscosity to superfluid density derived in the EFTs persists at finite temperature all the way to $T_C$.

The outline of the paper is as follows. In section 2, we review the main results of the effective field theory that governs the dynamics of the system at zero temperature. In section 3, we discuss the background geometry of the system and analyze the behavior of various background quantities. At low temperatures the system develops a new  scaling symmetry with nontrivial dynamical critical exponent $z$. In section 4, we calculate Hall and shear viscosities and conductivities. We conclude in section 5.

\section{Review of effective field theory predictions}
\label{sec:EFT}
In this section we review some of the results from the effective theory description of chiral superfluids that holds at zero temperature \cite{Golkar:2014paa}. We work in $2+1$ dimensions. Defining the dual gauge field $a_\mu$ as the hodge dual of the superfluid current $j= \frac 1{2\pi} *\!f$, the action up to first order in derivative expansion\footnote{In this power counting scheme $a_\mu$ is taken as $\mathcal O(p^{-1})$ because $n$ is $\mathcal O(p^0)$. The background fields $A_\mu$ and $g_{\mu\nu}$ are also taken as $\mathcal O(p^0)$.} is:
\begin{equation}
\label{eq:EFT_action}
S = \int\!d^3x\,\sqrt{-g} \left[
-\ce(n)+ \frac 1{4\pi} \varepsilon^{\mu\nu\rho}A_\mu f_{\nu\rho}
+\zeta(n) \epsilon^{\mu\nu\rho}u_\mu\d_\nu u_\rho
+ \kappa a_\mu J^\mu \right],
\end{equation}
where $A_\mu$ is the electromagnetic potential, the number density $n$ is defined as $n^2 = -j^\mu j_\mu=+f^2/8\pi^2$, $\ce(n)$ gives the energy density and $\zeta(n)$ is an undetermined function of the number density. Finally, the current $J^\mu$ is the Euler current\cite{Golkar:2014,Golkar:2014paa}:
\begin{equation}
\label{eq:Euler}
J^\mu =\frac{1}{8\pi} \varepsilon ^{\mu\nu\rho} \varepsilon^{\alpha\beta\gamma} u_\alpha
		\left( \nabla_\nu u_\beta \nabla_\rho u_\gamma 
		- \frac12 R_{\nu\rho\beta\gamma}\right),
\end{equation}
where $u^\mu$ is the normalized direction of the superfluid current $j^\mu = n u^\mu$. We note that the total charge associated to the Euler current on a spatial section is a half of the Euler characteristic of the surface \cite{Golkar:2014paa}.

It is important to note that since $a_\mu$ is the dual of the superfluid current, its charge counts the vortex number in the superfluid. By taking a variation of the  action \eqref{eq:EFT_action} with respect to $a_0$, we can see that there is a relationship between the vortex number, the total flux and the Euler characteristic of the space:
\begin{equation}
	\label{eq:topol_relation}
	N_V=N_\phi + \kappa\frac\chi2.
\end{equation}
This is in fact a constraint built into the effective action from topological considerations \cite{Read:1999fn,Golkar:2014paa} and in the case of a $p\pm ip$ superfluid, it can be summarized as the necessary relaionship between the flux and the topology such that the vector order parameter can be smoothly defined on the entire manifold. In terms of the effective theory parameters, this means:
\eq{
p\pm i p:~\kappa = \mp 2. 
}
Defining the superfluid shift $\mathcal S$ as the net flux
required to have no vortices when the superfluid is placed on a sphere, we see that\footnote{Comparing to \cite{Golkar:2014paa}, we have put the superfluid charge equal to 1.} $\mathcal S= \mp 2$ for $p\pm i p$.

Of particular interest are the predictions of this effective action with regards to the Hall conductivity and viscosity at zero frequency. For the $p+ip$ case,
\begin{align}
	\label{eq:Hall_visc}
	\eta_H=&-\frac{n}{2},\\
	\sigma_{xy}=&-\frac{n q^2 }{\mu^2}+\frac{2q^2\zeta(n)}{\mu^2}.
\end{align}
While the prediction for the Hall viscosity is universal, the Hall conductivity depends on the non-universal function $\zeta(n)$. Requiring the action to be Weyl invariant fixes the this function to be linear $\zeta(x)= \zeta_0 x$ and we get:
\begin{equation}
	\label{eq:Hall_cond}
	\sigma_{xy}=(2 \zeta_0 -1) \frac{n q^2}{\mu^2}.
\end{equation}
In what follows, we verify these predictions in the holographic setup and find that the Hall viscosity result is in fact robust at finite temperature.

\section{Holographic chiral superfluids}
\label{sec:background}
In this section we review the  standard  construction of holographic chiral superfluids \cite{Gubser:2008zu,Roberts:2008ns}. The bulk theory is  Einstein-Yang-Mills with group $SU(2)$,\footnote{Note that even though this holographic $p+ip$ superfluid is unstable to a rotation breaking $p_x$ phase \cite{Gubser:2008wv,Roberts:2008ns}, this is not an problem when calculating the response functions under consideration here. } where by turning on a chemical potential for a generator $\tau^3$ we explicitly break the global symmetry group of the boundary theory to $U(1)$. The W-bosons in the bulk provide us with natural charged vector operators in the boundary theory, which we use to construct a $p+ip$ order parameter spontaneously breaking the remaining $U(1)$. This spontaneous symmetry breaking is vector-like and breaks parity and time reversal.

\subsection{The superconducting phase}

Consider Einstein-Yang-Mills theory in asymptotically $AdS_4$ space:
\eq{
S = \int d^4x \sqrt{-g} \left[
\frac{1}{2\kappa^2}\left(R+\frac{6}{L^2}\right) - \frac{1}{4g^2}F^a_{\mu\nu} F^{a\mu\nu},
\right]
}
where $F^a_{\mu\nu} = \partial_\mu A_\nu^a - \partial_\nu A_\mu^a + f^{abc}A^b_\mu A^c_\nu,$ and we chose $G=SU(2),~f^{abc}=\epsilon^{abc}.$

Einstein's equations are simply
\eq{
\frac{1}{2\kappa^2} \left(R_{\mu\nu} -\frac{1}{2}R g_{\mu\nu} - \frac{3}{L^2} g_{\mu\nu} \right)=T_{\mu\nu},~T_{\mu\nu} = \frac{1}{4g^2} \left( F^a{}_{\mu\lambda} F^a{}_\nu{}^{\lambda} -\frac{1}{2}F^a{}_{\rho\sigma} F^a{}^{\rho\sigma}\right),
}
and the Yang-Mills equation is
\eq{
\nabla_\mu F^{a\mu\nu}+f^{abc} A^b{}_\sigma F^{c\sigma \nu}=0.
}
In what follows it will be advantageous to work with complex coordinates and generators,
\eq{
z=\frac{x+i y}{\sqrt 2},~\tau^\pm = \frac{\tau^1 \pm i \tau^2}{\sqrt 2}.
}

\begin{figure}[htbp]
\begin{center}
\includegraphics[scale=0.7]{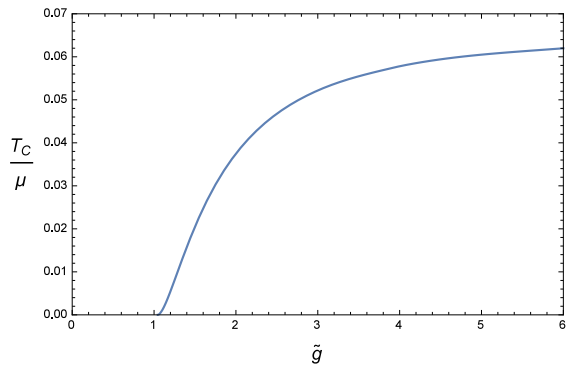}
\caption{Plot of the critical temperature as a function of $\gt$. }
\label{Tc_g}
\end{center}
\end{figure}

Our normal phase is the standard Reissner-Nordstrom black hole,
\eq{
ds^2 = -f dt^2 + \frac{dr^2}{f}+2r^2 dz d\zb ,~f=\frac{r^2}{L^2} - \frac{r_H(2g^2 r_H^2 + \k^2 \mu^2 L^2)}{2g^2L^2 r} + \frac{r_H^2 \k^2 \mu^2}{2g^2r^2},~ A = \mu (1-r_H/r) \tau^3 dt.
}
In this normal phase, the non-normalizable chemical potential explicitly breaks $SU(2)$ down to $U(1)$. At low enough temperature and large enough charge there may be a dynamical instability to developing vector hair. Consider the following ansatz: 
\eqn{
ds^2 &=& -h(r) dt^2 + 2 r^2 dz d\zb + \frac{dr^2}{f(r)},\nonumber \\
p_x+ip_y:~A &=& \phi(r) \tau^3 dt+w(r) (\tau^- dz + \tau^+ d\zb), \nonumber \\
p_x-ip_y:~A &=& \phi(r) \tau^3 dt+w(r) (\tau^+ dz + \tau^- d\zb).
\label{eq:flat_ansatz}
}
This is invariant under a combined rotation and gauge transformation
\eq{
p_x \pm i p_y: z \rightarrow e^{i\theta}z,~ U =e^{\pm i \theta \tau^3}.
}
The equations of motion for the background fields are independent of  chirality and reduce to
\eq{
f' + \frac f r - \frac{3r}{L^2} + \frac{\kappa^2}{g^2} \left( \frac{w^4}{2r^3} + \frac{w^2\phi^2}{r h} + f \left( \frac{ w'^2}{r}+\frac{r \phi'^2}{2h} \right) \right)=0
}
\eq{
h' + \frac{h}{r} - \frac{3r h}{L^2 f} + \frac{\kappa^2}{g^2}\left(
\frac{h w^4}{2r^3 f} - \frac{w^2 \phi^2}{rf} - \frac{h w'^2}{r} + \frac{r\phi'^2}{2}
\right)=0,
}
\eq{
\phi'' + \frac{2\phi'}{r} - \frac{2w^2 \phi}{r^2 f} - \frac{\kappa^2}{g^2} \left( \frac{w^2\phi^2}{r f h} + \frac{w'^2}{r} \right)\phi'=0
}
\eq{
w'' - \frac{w'}{r} + \frac{3rw'}{L^2 f} + \left(\frac{\phi^2}{fh} - \frac{w^2}{r^2 f} \right)w - \frac{\kappa^2}{g^2}\left(\frac{w^4}{2r^3 f} + \frac{r \phi'^2}{2h} \right) w'=0.
}
Note that our ansatz has two scaling symmetries \cite{Gubser:2008zu},
\eq{
r \rightarrow a r,~(t,x,y) \rightarrow (t,x,y)/a,~f \rightarrow a^2 f,~h \rightarrow a^2 h, ~w\rightarrow a w,~\phi \rightarrow a \phi,\nonumber}
\eq{
t \rightarrow b t,~h \rightarrow b^2 h,~ \phi \rightarrow b \phi.
\label{eq:bg_scaling}
}
By considering perturbations of $w$ on the Reissner-Nordstrom background and looking for normalizable solutions we can extract the critical temperature as a function of $\gt = gL/\k$, which we show in figure \ref{Tc_g}.
We find that there is a critical value of the coupling $\gt_C \approx 1.045$ where the critical temperature vanishes.

\begin{figure}
  \centering
	\begin{subfigure}[b]{0.46\textwidth}
			\centering
			\includegraphics[width=\textwidth]{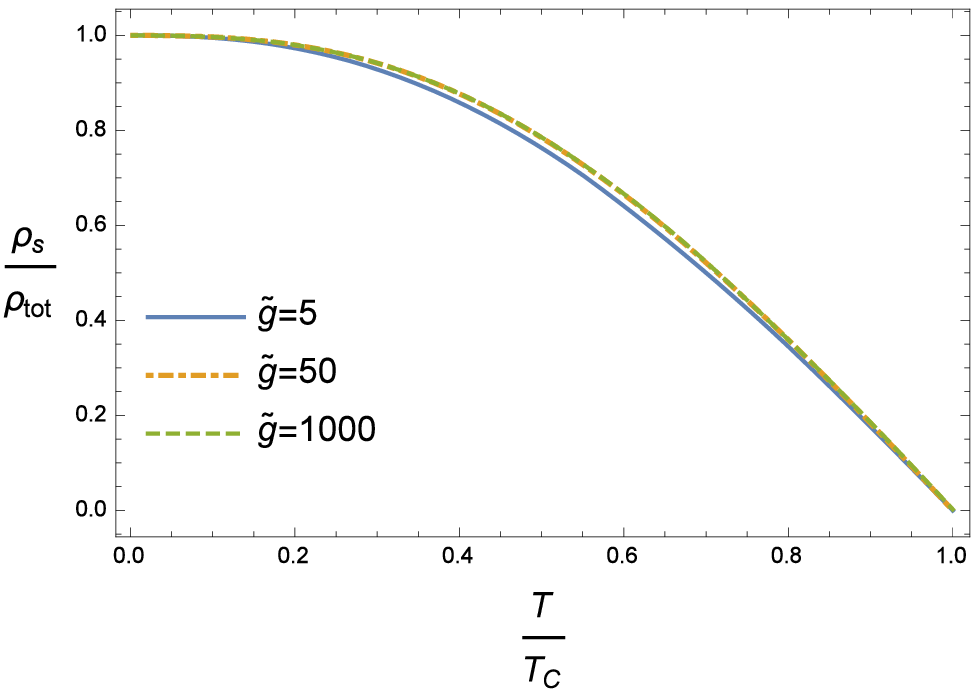}
			\label{fig:rho_s_O_rho_tot}
	\end{subfigure}
	\hspace{0.03\textwidth}
	\begin{subfigure}[b]{0.46\textwidth}
			\centering
			\includegraphics[width=\textwidth]{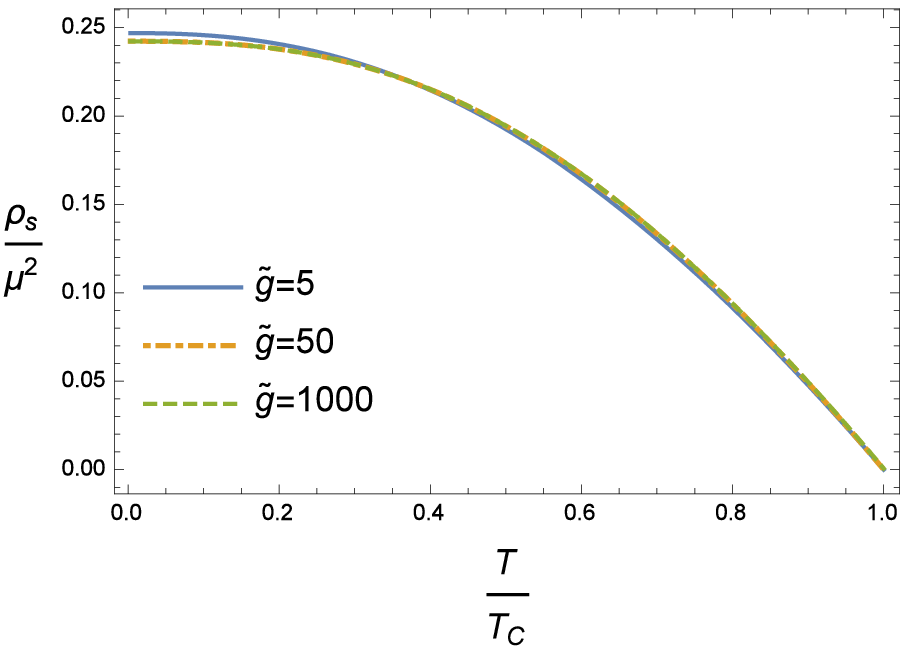} 
			\label{fig:rho_s_O_mu_sq}
	\end{subfigure} \vspace{-15pt}
  \caption{Plot of superfluid fraction and superfluid density (in units of  $\frac{L^2 }{g^2}$) vs temperature.}
  \label{fig:rhos}
\end{figure}

When considering the system at finite temperature we look for solutions with a regular horizon corresponding to $f$ vanishing linearly at $r=r_H$ and we will pick a gauge where $\phi(r_H)=0$ as well. Asymptotically the fields 	behave as
\eqn{
f = \frac{r^2}{L^2} - \frac M r + \ldots,~h = \frac{r^2}{L^2} - \frac M r+\ldots,~ f-h = \cO(r^{-2})
\nonumber}
\eq{
 w = w_0 + w_1/r + \ldots,~ \phi = \mu - \rho/r+\ldots\label{eq:asympt_flat_bg}
}
Following the standard holographic dictionary we interpret $\mu$ as the chemical potential and $w_0$ corresponds to an external source for the nonabelian current whose expectation value is
\eq{
\vev{J_{bdy}}=\frac{w_1}{g^2} (dz \tau^- +d\zb \tau^+).
}
To have spontaneous $U(1)$ breaking we consider only solutions with $w_0=0.$ The entropy, temperature, energy density and total charge density are
\eq{
S = \frac{2 \pi r_H^2}{\k^2},~ T = \frac{\sqrt{f'(r_H)h'(r_H)}}{4\pi},~\epsilon= \frac M {\k^2},~\rho_{total} = \frac{\rho}{g^2}.\label{eq:entropy_temp_general}
}
We can define the normal fluid charge density via the flux on the horizon and similarly the superfluid charge density as the difference between the total and the normal fluid charge \cite{Gubser:2008wv},
\eq{
\rho_n = \frac{r_H^2}{g^2} \phi'(r_H),~\rho_s = \rho_{total} - \rho_n.
}
Plots of the superfluid density  are given in figure \ref{fig:rhos}.  As expected, the ratio of superfluid to total charge  goes to one at zero temperature and vanishes linearly at $T_C$ as in mean field theory.

We should note that defining the superfluid density through the residue of the zero frequency pole of various correlators can be misleading. For example the $\sigma_{xx}$ pole only agrees with our definition at zero temperature, and even above $T_C$ the pole is nonzero as our system is exactly translationally invariant so there is infinite DC conductivity. While we strongly suspect that the definition we use for superfluid density will agree with the normalization of the Goldstone two-point function, we leave confirming this to future work.

\subsection{Low temperature and the Lifshitz throat}

\begin{figure}[htbp]
\begin{center}
\includegraphics[scale=0.8]{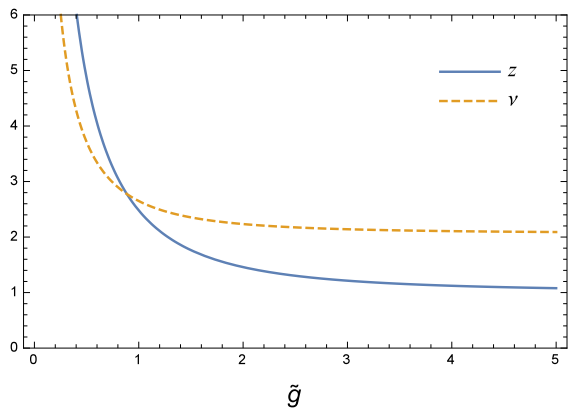}
\caption{Plot of the dynamical critical exponent $z$ and IR dimension $\Delta$ as a function of $\gt$. }
\label{exponents_g}
\end{center}
\end{figure}

At very low temperatures the geometry develops a Lifshitz throat. We can most easily see this after noticing that at low temperatures the solutions behaves very similarly to an s-wave holographic superconductor, where most of the electric charge in the solution is carried in the hair outside of the black hole and not on the horizon itself \cite{Horowitz:2009ij}. In the zero temperature limit, the degenerate horizon carries no charge and has vanishing area. In the extreme near-horizon limit, we expect a scaling solution with $\phi=0$. Examining the equations of motion with $\phi=0$ and again using the dimensionless coupling $\gt = gL/\k$ we find the solution
\eq{
f=\frac{6-\gt^2(z-1)^2}{2(2+z)}\frac{r^2}{L^2},~h=h_0 r^{2z},~w=\gt\sqrt{z-1}\frac rL, \label{eq:lifshitz_throat}
}
where $z$ is the real root of the cubic equation
\eq{
z^3 + z^2 + \frac{\gt^2-6}{\gt^2} z - \frac{3(2+\gt^2)}{\gt^2}=0.
}
and should not be confused with the complex coordinate on the plane.  Note that a positive root exists for any real $\gt$ and $z>1$. One can also check that the coefficient in front of $f$ in \eqref{eq:lifshitz_throat} is always positive. We plot $z(g)$ in figure \ref{exponents_g}. This throat has an anisotropic Lifshitz scaling symmetry \cite{Kachru:2008yh},
\eq{
t \rightarrow \lambda^z t,~ x \rightarrow \lambda x, r\rightarrow r/\lambda.
}

 From this Lifshitz throat we can turn on a perturbation in $\phi$, which (at zero frequency and momentum) can have scaling behavior
 \eq{
 \phi \sim a r^{-(1-z/2+\nu)}+b r^{-(1-z/2-\nu)},~\nu= \sqrt{ z(4+z)/4+3}.
 }
The $a$ mode diverges at small $r$ and vanishes at large $r$, while the $b$ mode vanishes at small $r$ and diverges at large $r$. We can therefore turn on a $b$-type perturbation and its backreaction will flow us to $AdS_4$ at large $r$.

\begin{figure}[ht]
  \centering
	\begin{subfigure}[b]{0.46\textwidth}
			\centering
			\includegraphics[width=\textwidth]{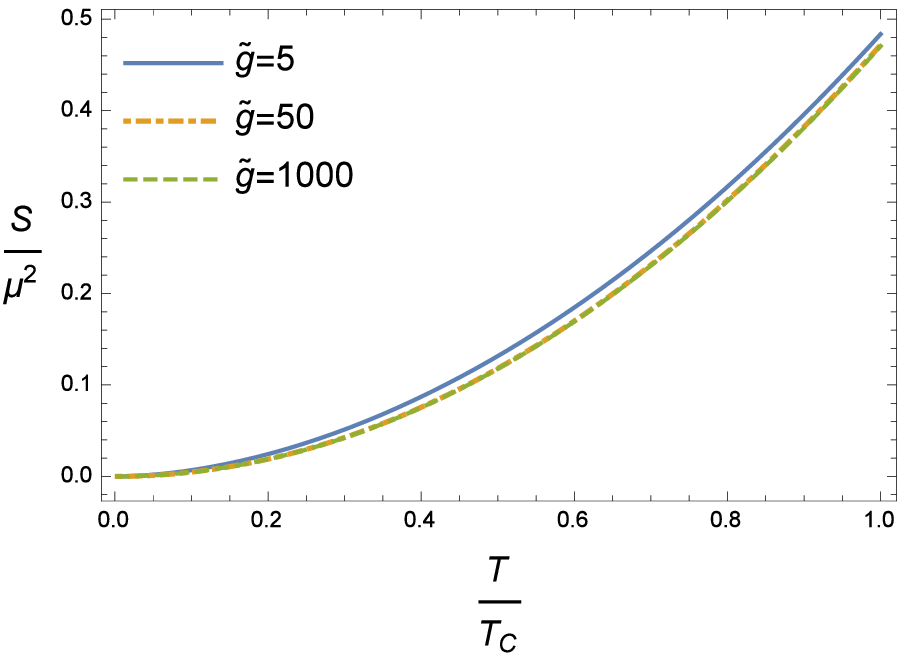}
			\label{fig:S}
	\end{subfigure}
	\hspace{0.03\textwidth}
	\begin{subfigure}[b]{0.46\textwidth}
			\centering
			\includegraphics[width=\textwidth]{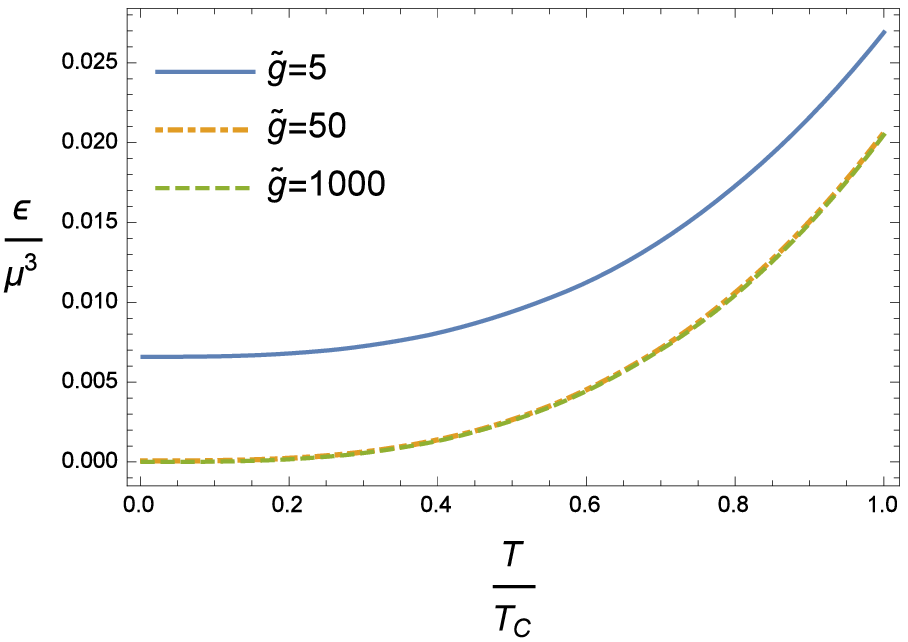} 
			\label{fig:M}
	\end{subfigure} \vspace{-15pt}
  \caption{Plots of entropy and energy density versus temperature respectively in units of $\frac{L^2}{\kappa^2}$ and energy density $\frac{L^4 }{\kappa^2}$.}
  \label{fig:S&M}
\end{figure}

We can also consider relevant perturbations in the other background field modes. There are normalizable static perturbations of $f,\chi,w$ which have the following behavior at large radius:
\eq{
\delta f = -\frac{M}{r^z},~\delta h = -\frac{h_0 L^2 (2+z^2)(3+2z+z^2)}{6z(2+z)}\frac{M}{r^{2-z}},~\delta w = \frac{L  \sqrt{3+2z(2+z)}}{2\sqrt{6}z\sqrt{1+z}}\frac{M}{r^{1+z}}.
}
Such an asymptotically Lifshitz solution has a mass density
\eq{
\epsilon = \frac{M L\sqrt h_0 \sqrt{3+2+z^2}}{\sqrt 6 \k^2}.
}
If we consider an asymptotically Lifshitz black hole at finite temperature, it must have $f \sim h \sim (r-r_H)$, and we can use the scaling symmetry \eqref{eq:bg_scaling} to argue that
\eq{
S \sim r_H^2,~ T \sim r_H^z,~ \epsilon \sim r_H^{2+z},
}
which can be repackaged into a more familiar form
\eq{
S \sim T^{2/z},~ \epsilon \sim T^{1+2/z},
}
as expected for a fixed point with dynamical critical exponent $z$. The numerical evaluation of these quantities is given in plot \ref{fig:S&M} and matches this scaling. We note that due to numerical difficulty, the smallest value of $\tilde g$ that we considered was $\tilde g=5$.

\subsection{Superfluid on a sphere}
\label{sec:sphere}

The superfluid shift, defined in \cite{Golkar:2014paa} for relativistic superfluids in 2+1 dimensions, is the net flux required to have no vortices when the superfluid is placed on a sphere. For a chiral p-wave superfluid with charge equal to 1, the shift is related to the chirality:
\eq{p_x\pm ip_y:~\mathcal S = \mp 2. \label{eq:chiral_shift_pred}
}

As the holographic model under consideration is meant to describe a $p\pm ip$ superfluid, we may ask whether or not it can reproduce this result. The answer it that it indeed does. First, as the arguments for this constraint are purely topological they would straight-forwardly apply to the holographic setup as well. However,  putting this system on a sphere, one finds the constraint is also embedded in the bulk equations of motion. To see this, we look  for vortex-free solutions in asymptotically \emph{global} AdS. Parts of the calculation are simpler in projective coordinates, where the sphere metric is given by:
\eq{
d\Omega^2 = d\theta^2 + \sin^2\theta d\phi^2 = \frac{2 dz d\zb}{\left(1+z\zb/2 \right)^{2}},~z=\sqrt{2} e^{i\phi} \tan\left[\theta/2 \right].
}

We also need to turn on a constant magnetic flux through the sphere given (in the upper hemisphere) by:
\eq{
A \supset \frac{N_\phi}2(1-\cos\theta)d\phi \tau^3 = i \frac{N_\phi}2 \frac{z d\zb - \zb dz}{2+z\zb}\tau^3.
}
We generalize the ansatz \eqref{eq:flat_ansatz} for the sphere such that it retains the same rotation properties about the north pole as flat space:
\eq{
p\pm ip:~A \supset w(r)\left( \tau^\mp dz + \tau^\pm d\zb\right)H(z \zb).
}
The full ansatz for vortex free solutions is
\eqn{
ds^2 &=& -h(r) dt^2 + r^2 d\Omega^2 + \frac{dr^2}{f(r)},\nonumber \\ 
A &=& \tau^3 \left(\phi(r)  dt +  i \frac{N_\phi}2 \frac{z d\zb - \zb dz}{2+ z\zb} \right) + \left( \tau^\pm dz + \tau^\mp d\zb\right)H(z \zb) w(r).
}
Plugging this into the equations of motion, we find that the bulk Yang-Mills equations constrain both $N_\phi$ and $H$. First, looking at the $(\ldots)^{\pm~r}$ component, we find the constraint
\eq{
H'(z\zb) = \pm \frac{N_\phi}2\frac{H(z \zb)}{2+z \zb},
}
and from the $(\ldots)^{3~z,~3~\zb}$ component we find
\eq{
H'(z \zb) = \frac{\pm N_\phi-4}{6}\frac{H(z \zb)}{2+z \zb},
}
which gives the nontrivial constraint 
\eq{
N_\phi = \mp 2,}
verifying our prediction from more general EFTs \eqref{eq:chiral_shift_pred}, as well as a functional form for $H$,
\eq{
H = \frac{1}{1+z \zb/2}.
}
 In terms of spherical coordinates and standard generators, the ansatz is (again only valid in the upper patch):
 \eqn{
 A &=& \tau^3 (\phi(r) dt \pm(1-\cos\theta)d\phi)  \nonumber \\
 &+&
w(r) \left[\left(\cos\phi d\theta -\sin\theta\sin\phi d\phi \right)\tau^1
\mp
\left(\sin\phi d\theta +\sin\theta\cos\phi d\phi \right)\tau^2
 \right] 
 }
Viewed from above, this corresponds to an order parameter configuration which is pointing in the same direction across the entire patch.

With this ansatz one can work out the background equations of motion as a set of ODEs. They are similar to those in planar coordinates with additional terms coming from the radial slicing curvature and the magnetic field. Solutions with spontaneous symmetry breaking still exist below a critical temperature and at large enough chemical potential. Note that due to the magnetic field at low temperatures on the sphere the system will develop a $AdS_2 \times S^2$ near horizon geometry instead of a Lifshitz throat.

\section{Transport}
\label{sec:perturbations}
To study transport properties  we look at perturbations on top of the background discussed in the previous section. We will only consider finite frequency but zero spatial momentum perturbations of the form $\delta\phi(r)\, e^{-i \omega t}$. This will allow us to separate the scalar, vector and tensor sectors following the mode decomposition of \cite{Son:2013xra}. For the sake of brevity, we restrict to the case of $p+ip$. Since time reversal switches the chirality of the VEV, the results for $p-ip$ can be readily derived from what follows by taking $\omega \to -\omega$ and taking the complex conjugate of the fields. 

\subsection{Viscosity and tensor perturbations}
\label{subsec:tensor}
In this section we calculate the shear and Hall viscosity of our system (conformal symmetry causes the bulk viscosity to vanish).
To caclulate the stress viscosity transport coefficients, we look at metric perturbations which give us boundary stress response \cite{Balasubramanian:1999re}. More explicitly, metric fluctuations behave asymptotically as
\eq{\label{eq:g_exp}
\delta g_{\mu\nu} =r^2 \gamma_{\mu\nu}e^{-i\w t}+\ldots + h_{\mu\nu}e^{-i\w t}/r+\ldots
}
Here  we have picked the gauge $\delta g_{rr}=\delta g_{r\mu}=0,$ and so $\mu$ runs over $t,x,y$. When $\gamma_{\rho\sigma}$ is the only nonnormalizable component that is nonzero we have the stress response,
\eq{\label{eq:TT_gen}
\vev{T_{\mu\nu}(-\w)T_{\rho\sigma}(\w)} = -\frac{3}{2\k^2 L^2} \frac{h_{\mu\nu}}{\gamma_{\rho\sigma}}.
}
To measure the Hall and shear viscosities, consider a tensor perturbation where the only  nonnormalizable perturbation turned on is $\gamma_{xy}.$ 
\begin{align}
\vev{T_{xx}(-\w)T_{xy}(\w)} =& -i\w \eta_H = -\frac{3}{2\k^2 L^2} \frac{h_{xx}}{\gamma_{xy}},\\
\vev{T_{xy}(-\w)T_{xy}(\w)} =& -i\w \eta_{sh} = -\frac{3}{2\k^2 L^2} \frac{h_{xy}}{\gamma_{xy}}.
\label{eq:TT_H_SH}
\end{align}
How do such perturbations behave in the superfluid phase? The tensor modes on the $p_x + i p_y$ background are 
\eq{
\delta g_{xy} := a_1,~\delta A_y^1 = + \delta A^2_x := a_2,~\delta g_{xx} = -\delta g_{yy} := a_3,~\delta A^1_x = - \delta A^2_y := a_4.
}
The linearized equations take the form
\eq{
\partial_r^2\vec{a} + K\cdot \partial_r\vec{a} + M \cdot \vec{a} = 0,
}
where
\eq{
K = \left(\begin{array}{cccc}
K_1 & \frac{4\k^2 w'}{g^2} & 0 & 0 \\
-\frac{w'}{r^2} & K_2 & 0 & 0 \\
0 & 0 & K_1 & \frac{4\k^2 w'}{g^2}  \\
0 & 0 & -\frac{w'}{r^2} & K_2
\end{array}\right)
,~
M = 
\left(\begin{array}{cccc}
M_1 & -\frac{4 \k^2 w \phi^2}{g^2 fh} & 0 & - \frac{4i \k^2 \w w \phi}{g^2 fh} \\
M_2 & M_3 & - \frac{i\w w\phi}{r^2 fh} &  \frac{2i\w \phi}{fh} \\
0 & \frac{4i \k^2 \w w \phi}{g^2 fh} & M_1 & -\frac{4 \k^2 w \phi^2}{g^2 fh} \\
\frac{i\w w\phi}{r^2 fh}  & - \frac{2i\w \phi}{fh} & M_2 & M_3
\end{array}\right)
}
\eqn{
K_1&=&-\frac 3r + \frac {3r}{L^2 f}-\frac{\k^2 w^4}{2g^2r^3f}-\frac{r \k^2\phi'^2}{2g^2h},~K_2=-\frac 1r + \frac{3r}{L^2f}-\frac{\k^2w^4}{2g^2r^3f} - \frac{r\k^2 \phi'^2}{2g^2 h} , \nonumber \\
M_1 &=& \frac{4}{r^2} - \frac{6}{L^2 f}+\frac{\w^2}{fh} + \frac{\k^2 w^4}{g^2 r^4 f} + \frac{2\k^2 w^2 \phi^2}{g^2 r^2 f h}-\frac{2\k^2 w'^2}{g^2 r^2} + \frac{\k^2 \phi'^2}{g^2 h}, \nonumber \\
M_2 &=& -\frac{w^3}{r^4 f}+\frac{2w'}{r^3},~M_3 = \frac{\w^2}{fh}+\frac{w^2}{r^2 f}+\frac{\phi^2}{fh}.
}
Consider the solution which behaves asymptotically as:
\eqn{
a_1 = a_{10} r^2 + \frac12 L^4 \w^2 a_{10} + \frac{a_{13}}{r}+\ldots,~a_2 = \frac{a_{21}}{r}+\ldots,~a_3 = \frac{a_{33}}{r}+\ldots,~a_4 = \frac{a_{41}}{r}+\ldots}
The  final formulae for the viscosities are
\begin{align}
	\eta_H(w)=&\frac{-3 i}{2\w\k^2 L^2} \frac{a_{33}}{a_{10}},\\
\eta_{SH}(\w)=&\frac{-3 i}{2\w\k^2 L^2} \frac{a_{13}}{a_{10}}.
\label{eq:etas}
\end{align}

\subsubsection{Hall viscosity}

\begin{figure}
  \centering
	\begin{subfigure}[b]{0.4\textwidth}
			\centering
			\includegraphics[width=\textwidth]{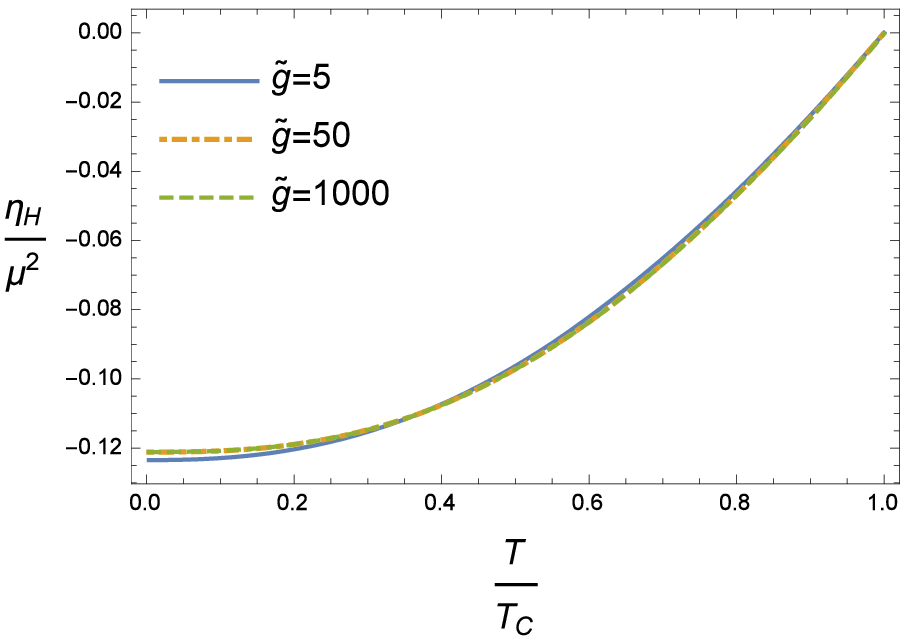}
			\label{fig:eta_H_musq}
	\end{subfigure}
	\hspace{0.03\textwidth}
	\begin{subfigure}[b]{0.425\textwidth}
			\centering
			\includegraphics[width=\textwidth]{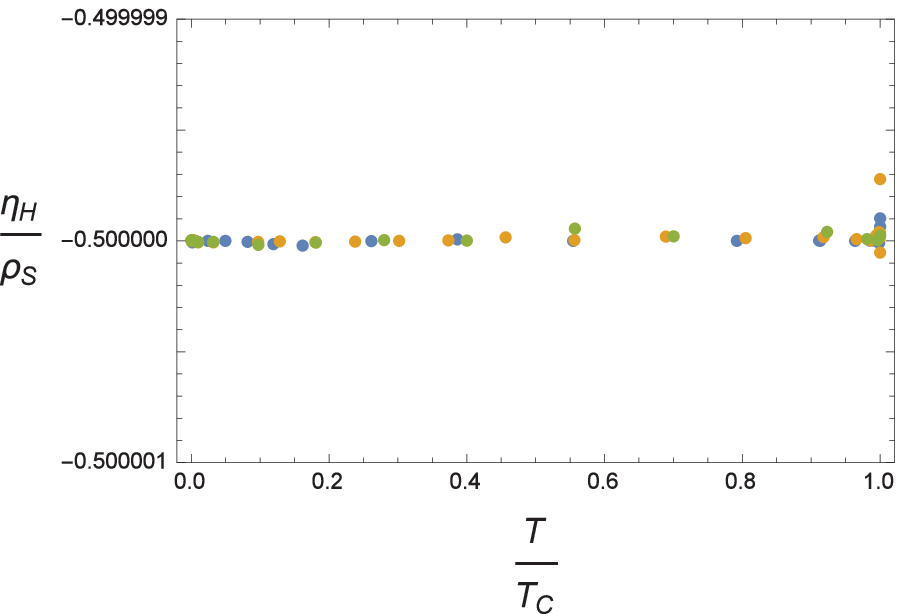} 
			\label{fig:eta_H_rhos}
	\end{subfigure} 
  \caption{(a) Plot of $\frac{\eta_H}{\mu^2}$. (b)  Plot of $\frac{\eta_H}{\rho_S}$.}
  \label{fig:eta_H}
\end{figure}

We first look at the Hall viscosity, which we compute by numerically integrating the equations of motion and using the formulae \eqref{eq:etas}. The Hall viscosity vs temperature results are given in plot \ref{fig:eta_H}a. We see that its magnitude increases as we approach zero temperature. 
Normalizing Hall viscosity value by the superfluid density (figure \ref{fig:eta_H}b), we see that the ratio equal to $-1/2$ ($\pm 10^{-7}$, within the numerical error we work at) over the entire temperature range and for all values of the coupling constant. This is the behavior that was predicted at zero tempereature from an effective field theory \eqref{eq:Hall_visc}. Here we see that its validity is more general and implies that at finite temperature, there is no contribution to Hall viscosity coming from the normal fluid sector. This is not surprising, as our normal fluid does not break parity (see \cite{Liu:2014gto} for a holographic model of a normal relativistic fluid which exhibits Hall viscosity). We do not have an analytic proof that $\eta_H/\rho_S = -1/2$ but suspect that it should be possible, likely in a manner similar to the proof of the shear viscosity bound \cite{Kovtun:2004de}.

A possible way of understanding this ratio is to note that the ratio of the Hall viscosity to the ``orbital angular momentum per particle'' has been shown to be equal to 1/2 in gapped or topological phases at zero temperature \cite{ReadRezayi:2010}. However, if we consider the fact that the vector-like order parameter carries angular momentum equal to one, then the superfluid density would be equal to the orbital angular momentum per particle and our numerical result can be considered as  a generalization of the viscosity-angular momentum relationship.

\subsubsection{Shear viscosity}
\begin{figure}
  \centering
	\begin{subfigure}[b]{0.46\textwidth}
			\centering
			\includegraphics[width=\textwidth]{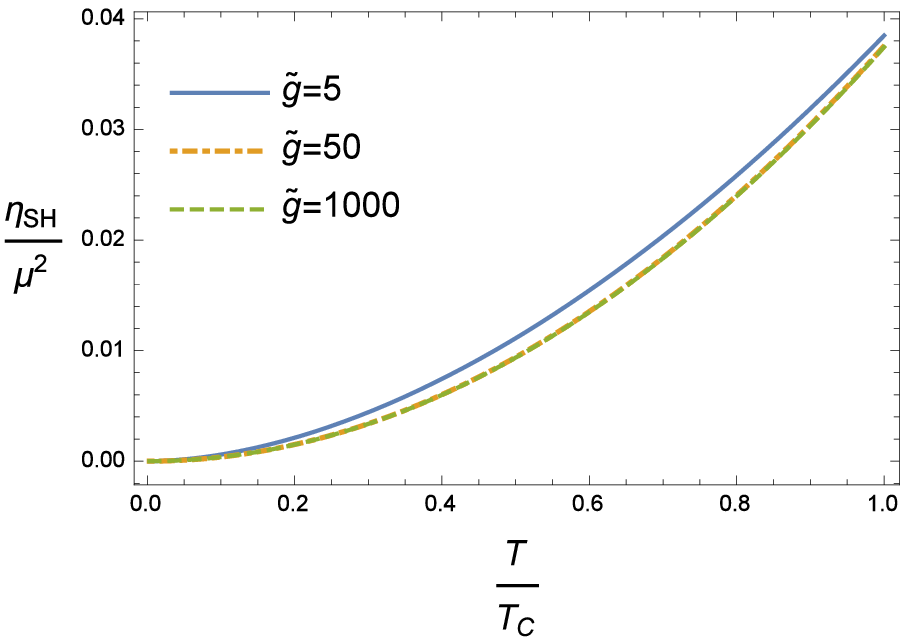}
			\label{fig:eta_S_5}
	\end{subfigure}
	\hspace{0.01\textwidth}
	\begin{subfigure}[b]{0.48\textwidth}
			\centering
			\includegraphics[width=\textwidth]{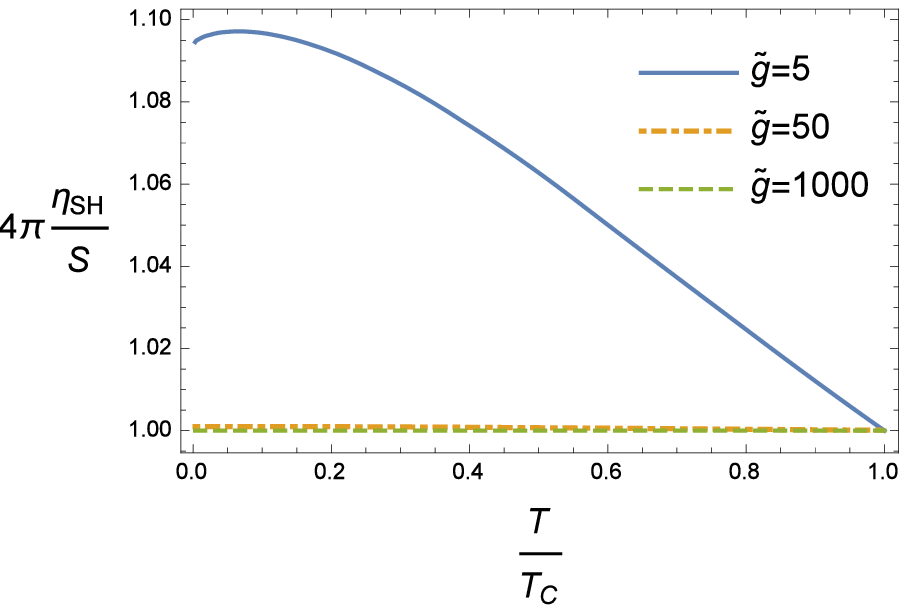} 
			\label{fig:eta_S_Norm}
	\end{subfigure} 
  \caption{(a) Plot of shear viscosity normalized by the chemical potential. (b)  Plot of $\frac{4\pi\eta_{SH}}{S}$. Note that the dip near zero temperature is also present at other values of $\gt$.}
  \label{fig:eta_S}
\end{figure}

The plots for shear viscosity vs. temperature at various values of the coupling constant can be seen in figure \ref{fig:eta_S}a. As expected this dissipative effect vanishes in the zero temperature limit. Near $T_C$,  where the background geometry is Reissner-Nordstrom, we know that the ratio $\eta_{sh}/S=1/4\pi$. The plot of this ratio at other temperatures is given in figure \ref{fig:eta_S}b. We see that as the temperature is reduced, this ratio increases, satisfying the bound \cite{Kovtun:2004de}. However, very close to the zero temperature limit the ratio starts to decrease again. We do not understand the origin of this dip. However, we can deduce that it must be coming from subleading terms as the shear viscosity and entropy must have the same scaling behavior in the IR Lifshitz fixed point.

\subsection{Conductivity and vector perturbations}
\label{subsec:vector_pert}

We consider the six following vector modes with harmonic time dependence:
\eq{
\delta g_{tx},~\delta g_{ty},~\delta A^3_x,~\delta A^3_y,~\delta A^1_t,~\delta A^2_t.
}
where we have gauged away the other vector modes. We are interested in calculating  the electric conductivity, defined by
\eq{
J_i = \sigma_{ij}E^j.
}
It is useful to work in the chiral/antichiral basis as these modes decouple from eachother.
\eq{
z=\frac{x+iy}{\sqrt2},~\tau^\pm = \frac{\tau^1 \pm i \tau^2}{\sqrt2},\nonumber }
This allows us to directly compute the eigenvalues of the conductivity matrix,
\eq{
\sigma_z = \sigma_{xx}+i\sigma_{xy},~\sigma_\zb = \sigma_{xx} - i \sigma_{xy}.
}
Our bulk modes are similarly organized as
\eq{
~\delta A^3_z=\frac{\delta A^3_x+i\delta A^3_y}{\sqrt{2}},~\delta A^\pm_t = \frac{\delta A^1_t\pm i A^2_t}{\sqrt{2}},~\delta g_{tz}=\frac{\delta g_{tx}+i\delta g_{ty}}{\sqrt2}.
}
Defining
\eq{
\vec{a}=\left( 
\delta A^3_z,~\delta A^ + _t,~\delta g_{tz}\right)\;\; \vec{\bar a}=\left( \delta A^3_\zb,~\delta A^-_t,~\delta g_{t\zb}
\right),\label{eq:vector_fields}
}
the $\vec a$ and $\vec{\bar a}$ modes are completely decoupled. In what follows, we restrict the discussion to the $\vec a$ modes. The linearized equations for the $\vec a$ modes are given by\footnote{The $\vec{\bar a}$ mode equations are simply derived by a time reversal transformation which takes  $\omega \to -\omega$ and  conjugates the fields (i.e.  it takes $a \to \bar a$).}:
\eq{
 \partial_r^2 a_1+
K\cdot \partial_r \vec{a}+
M\cdot  \vec{a}=0,
}
where 
\eq{
K=\left(\begin{array}{ccc}
k & 0 & 0  \\
\frac{hw}{r^2(\phi - \w)} & 1 & 0  \\
-\frac{2\k^2 h w^2}{g^2r^2(\phi - \w)} & 0 & 1 
\end{array}\right),\;
M=\left(\begin{array}{ccc}
N & -\frac{w}{h} m& \frac{w^2}{r^2h} m  \\
-\frac{h w'}{r^2(\phi-\w)} & -\frac{\phi'}{\phi-\w} & \frac{w\phi'}{r^2(\phi-\w)}-\frac{w'}{r^2} \\
\frac{2\k^2h w w'}{g^2r^2(\phi-\w)} + \frac{2\k^2\phi'}{g^2}& \frac{2\k^2 w \phi'}{\phi - \w} & -\frac 2r - \frac{2\k^2 w^2\phi'}{g^2r^2(\phi-\w)} 
\end{array}\right)
\notag}
where
\eq{
k=\frac{2\k^2 w^2 \phi'}{g^2 r^2 (\phi - \w)} +\frac{3r}{L^2 f} - \frac 1r - \frac{r \k^2 \phi'^2}{2g^2 h}\notag,
}
\eq{
N=\frac{\w^2}{fh}-\frac{2\k^2 w w' \phi'}{g^2r^2(\phi-\w)-\frac{w^2}r^2f}-\frac{2\k^2\phi'^2}{g^2h},
~m= \left( \frac{2\k^2\phi'^2}{g^2  (\phi - \w)}+ \frac{\phi + \w}{f} \right).
\notag}

At large radius, the fields \eqref{eq:vector_fields} will be of the form
\eq{
a_1 = a_{10}+a_{11}/r+\ldots, a_2 = a_{20}+\frac{a_{30}w_1}{r}-\frac{a_{11}w_0a_{10}w_1}{L^2(\mu-\w)r}-\frac{(a_{20}-a_{30}w_0)\rho}{(\mu-\w)r}+\ldots,\nonumber
}
\eq{
a_3 = a_{30}r^2 +\left(a_{10}\rho+\frac{w_0(a_{11}w_0-a_{10}w_1)}{L^2(\mu-\w)}+\frac{w_0\rho(a_{20}-a_{30}w_0)}{\mu-\w} \right) \frac{2\k^2}{3g^2 r}+\ldots
}
It is important with these perturbations do not only turn on a linearized non-normalizable $ A^3_{z,\zb}$ by way of $a_{10}$ but also possibly  $ A^+_t$ and $g_{tz}$. These  may be removed by a linearized gauge transformation \cite{Roberts:2008ns} and diffeomorphism and do not change the equations of motion. However, they affect the formulae for conductivity. Consider a linearized gauge transformation followed by a diffeomorphism
\eq{
\lambda = e^{-i\w t}\lambda^+ \tau^-,~\xi = e^{-i\w t} \xi^z \partial_\zb  ,
}
which we fix by requiring after the gauge transformation $A^+_t,~g_{tz}\sim \cO(1/r)$. This is done via
\eq{
\xi^z = -\frac{ia_{30}}\w,~\lambda^+ =  \frac i{2}\frac{(a_{20}-2a_{30}w_0)}{\mu-\w}.
}
After this fixing we find the conductivity is\footnote{It is clear from this formula that there will be a pole in the imaginary part of the conductivity at $\omega = \mu$. This is isospin precession which is undamped and temperature independent and follows directly from the nonabelian Ward identities \cite{Iqbal:2010eh,chrisprivate}.}
\eq{
\frac{\sigma_z}{\sigma_n} =  \frac{1}{i\w}\frac{a_{11}(\mu - \w)+(a_{20}-2 a_{30}w_0)w_1}{a_{10}(\mu - \w)+w_0(a_{20}-2 a_{30}w_0)},
}
where we have normalized by the value of the conductivity at $\omega \rightarrow \infty$,
\eq{
\sigma_n = \sigma_{xx}(\w \rightarrow \infty) = \frac{L^2}{g^2}.
}

Since we are interested in backgrounds with spontaneous symmetry breaking where $w_0=0$, these formulae simplify and we have:
\eq{
\frac{\sigma_z}{\sigma_n}=\frac{a_{11}}{i \w a_{10}} + \frac{ w_1 a_{20}}{i \w a_{10}(\mu - \w)}.
}
The conductivity for the anti-chiral modes is derived from this by complex conjugation and taking $\omega \to -\omega$:
\eq{
\frac{\sigma_{\bar{z}}}{\sigma_n}=\frac{\bar a_{11}}{i \w \bar a_{10}} + \frac{ w_1 \bar a_{20}}{i \w \bar a_{10}(\mu + \w)}.
}

\begin{figure}[ht]
  \centering
	\begin{subfigure}[b]{0.46\textwidth}
			\centering
			\includegraphics[width=\textwidth]{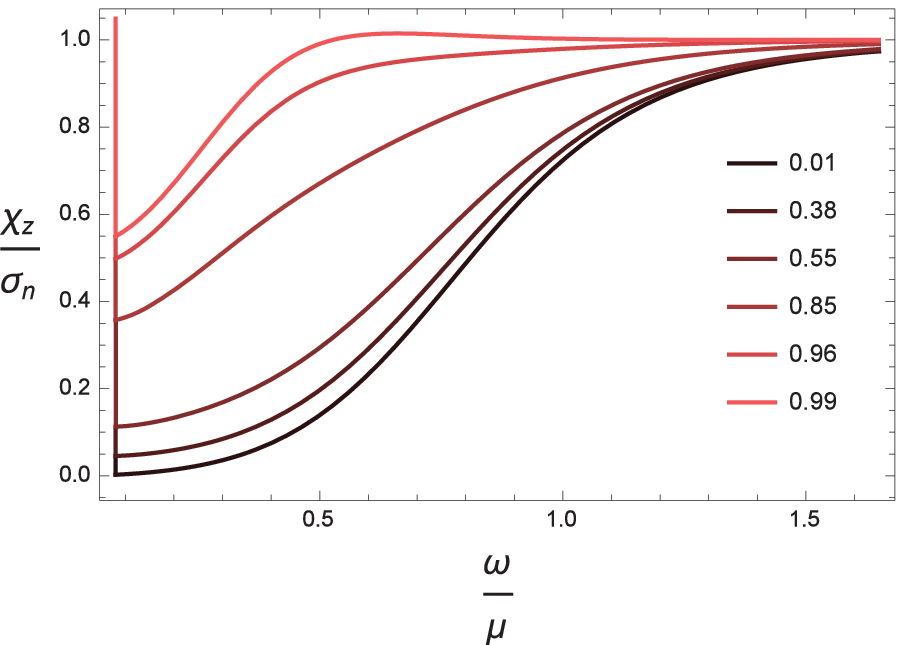}
			\label{fig:chi_z}
	\end{subfigure}
	\hspace{0.04\textwidth}
	\begin{subfigure}[b]{0.46\textwidth}
			\centering
			\includegraphics[width=\textwidth]{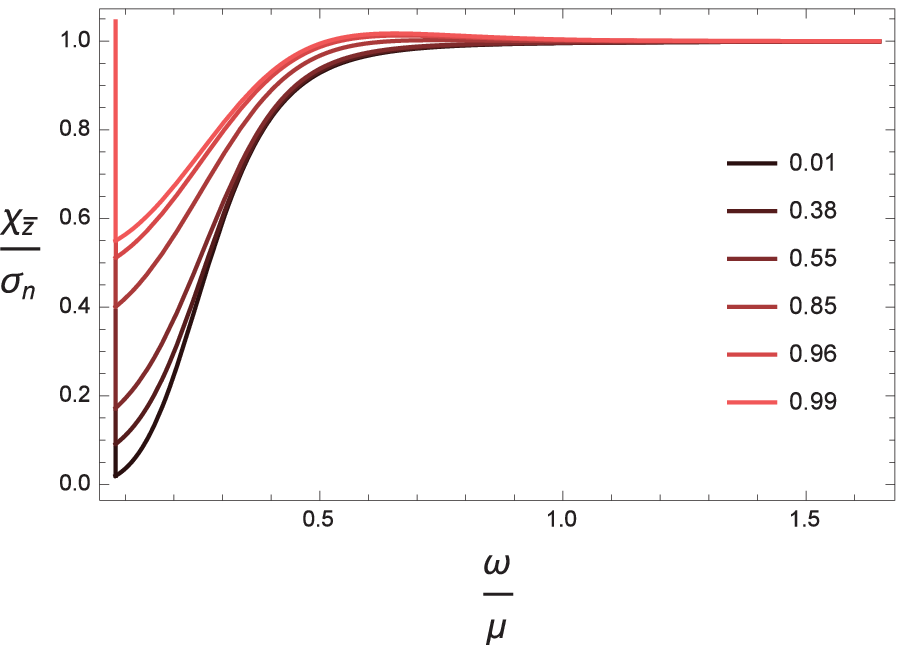} 
			\label{fig:chi_zb}
	\end{subfigure}\vspace{-15pt}
  \caption{Plots of $\chi_z$ (left) and $\chi_{\bar z}$ (right) for  various values of  $T/T_C$.}
  \label{fig:spectral}
\end{figure}

\begin{figure}\vspace{15pt}
  \centering
	\begin{subfigure}[b]{0.46\textwidth}
			\centering
			\includegraphics[width=\textwidth]{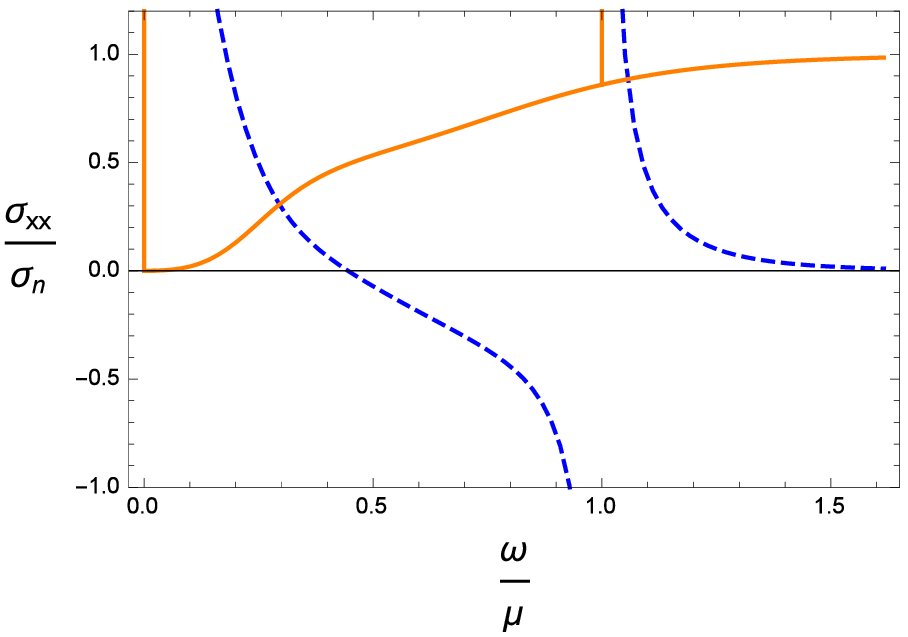}
			\label{fig:sigma_xx}
	\end{subfigure}
	\hspace{0.04\textwidth}
	\begin{subfigure}[b]{0.46\textwidth}
			\centering
			\includegraphics[width=\textwidth]{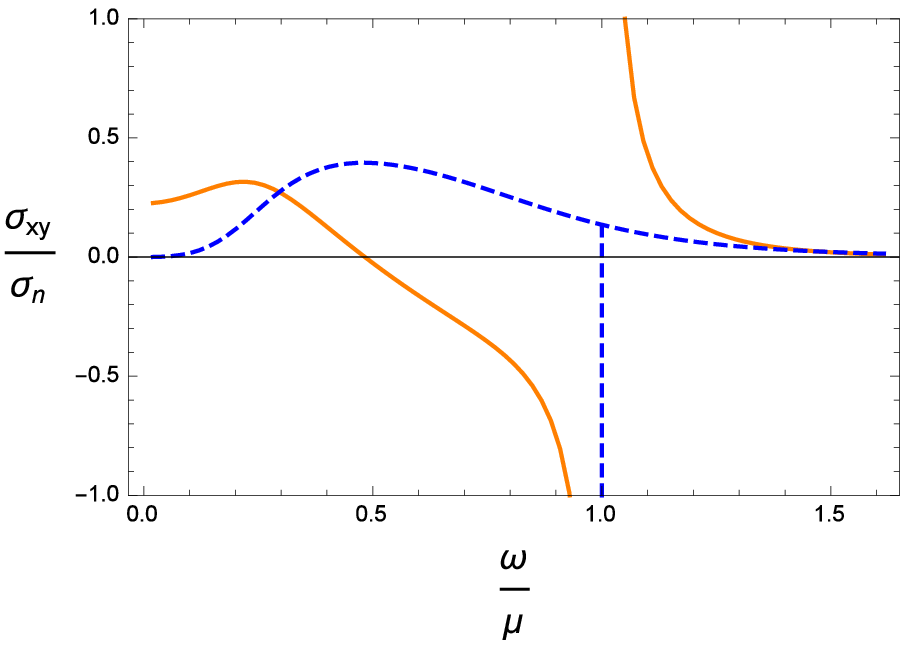} 
			\label{fig:sigma_xy}
	\end{subfigure} \vspace{-15pt}
  \caption{Plots of conductivities at zero temperature. The orange solid line is the real part and the blue dashed line denotes the imaginary part. }
  \label{fig:cond}
\end{figure}

The typical plots for the spectral densities can be seen in figure \ref{fig:spectral}. We note that these are qualitatively the same as their strict probe counterparts\footnote{The one exception is the dip in the spectral density near zero frequency at $T_C$. In the probe limit the dip vanishes.} in \cite{Roberts:2008ns}. We see that there is a pseudogap, i.e. a depletion of states as we approach zero temperature, which is a characteristic of holographic superconductors. The respective plots for the conductivities can be seen in figure \ref{fig:cond}. Near $T_C$, where the order parameter vanishes, parity is restored, which implies $\chi_z$ and $\chi_{\bar{z}}$ are equal. This implies that the DC conductivity must vanish.

  By comparing the values of $\sigma_{xy}$ to the superfluid density in the zero temperature limit, we can read off the coefficient of the non-universal $\zeta$ term in the effective field theory (eq. \eqref{eq:Hall_cond}). In the probe limit, this ratio is exactly equal to one. Turning on the back reaction of the geometry, we see that for large values of $\tilde g$,  the deviation from probe goes as ${\tilde g}^{-2}$. A fit to the ratio for various $\tilde g$ reveals that $\zeta_0=1-1.15\, {\tilde g}^{-2}+1.05\, {\tilde g}^{-4} + \mathcal O( {\tilde g}^{-6})$. As expected, this coefficient is non-universal and depends on the details of the theory and can also vary by temperature. We suspect that this is the reason why unlike the ratio of Hall viscosity to superfluid density which is independent of temperature, the same cannot be said for Hall conductivity.

\section{Conclusions and outlook}
\label{sec:conc}
In this paper we analyzed a chiral superfluid model and computed response functions  away from the probe limit. While some characteristics of the system (e.g. shape of electric conductivity plots) remain the same, there are some quantities that can only be calculated away from the probe limit. Due to the emergent Lifshitz geometry that governs the infrared dynamics, many of the thermodynamic variables are quantitatively different.

Looking at the viscosity response of the system, we found that the Hall viscosity to superfluid density ratio matches the zero temperature EFT prediction even at finite temperature, as does the superfluid shift. We also computed  the shear viscosity and found that the viscosity-entropy ratio respects the holographic bound, but has a feature near zero temperature where it decreases. Our results for transport are numerical only and it would be interesting to understand them from a first principle approach.

Unlike the low energy EFT analysis which only kept track of the Goldstone degree of freedom, this holographic model has the capacity to describe vortices. While these have been studied in the probe limit near critical magnetic fields \cite{Murray:2011gr} it would be very interesting to study them with back reaction. It would also be interesting to study the structure of bound states that arise on different topologies either due to the vortices created from a mismatch in the magnetic flux or in the presence of boundaries. 


\section*{Acknowledgments}
It is a pleasure to thank Kathryn Levin, Eun-Gook Moon and  Dam T. Son for insightful discussions.  This work is supported, in part, by DOE grant DE-FG02-13ER41958. S.G. is supported in part by NSF MRSEC grant DMR-1420709.

\addcontentsline{toc}{section}{Bibliography}
\bibliographystyle{JHEP}
\bibliography{HoloSF}

\end{document}